\documentclass[sigconf,authorversion]{acmart}

\def\BibTeX{{\rm B\kern-.05em{\sc i\kern-.025em b}\kern-.08emT\kern-.1667em\lower.7ex\hbox{E}\kern-.125emX}}

\usepackage{listings} 
\usepackage{caption}
\usepackage{subcaption}

\usepackage{placeins} 

\usepackage[nolist]{acronym}
\begin{acronym}
\acro{AOSP}{Android Open Source Project}
\acro{ACL}{Asynchronous Connection-Less}
\acro{ABI}{Application Binary Interface}
\acro{AUX}{auxiliary}
\acro{AWS}{Amazon Web Services}
\acro{JVM}{Java Virtual Machine}
\acro{ADB}{Android Debug Bridge}
\acro{HCI}{Host Controller Interface}
\acro{RAM}{Random Access Memory}
\acro{ROM}{Read Only Memory}
\acro{IoT}{Internet of Things}
\acro{LNA}{Low Noise Amplifier}
\acro{AGC}{Automatic Gain Control}
\acro{UART}{Universal Asynchronous Receiver-Transmitter}
\acro{TLV}{Type-Length-Value}
\acro{EEPROM}{Electrically Erasable Programmable Read-Only Memory}
\acro{UI}{User Interface}
\acro{RNG}{Random Number Generator}
\acro{NIST}{National Institute for Standards and Technology}
\acro{IO}{Input-Output}
\acro{LM}{Link Manager}
\acro{LMP}{Link Manager Protocol}
\acro{LCP}{Link Control Protocol}
\acro{LL}{Link Layer}
\acro{WEP}{Wired Equivalent Privacy}
\acro{SIG}{Special Interest Group}
\acro{BR}{Basic Rate}
\acro{EDR}{Enhanced Data Rate}
\acro{LE}{Low Energy}
\acro{BLE}{Bluetooth Low Energy}
\acro{TDD}{Time-Division Duplex}
\acro{AFH}{Adaptive Frequency Hopping}
\acro{L2CAP}{Logical Link Control and Adaption Protocol}
\acro{PDU}{Program Data Unit}
\acro{USB}{Universal Serial Bus}
\acro{RFCOMM}{Radio Frequency Communications}
\acro{SDP}{Service Discovery Protocol}
\acro{TCP}{Transmission Control Protocol}
\acro{QoS}{Quality of Service}
\acro{HEC}{Header-Error-Check}
\acro{LLID}{Logical Link ID}
\acro{CRC}{Cyclic Redundancy Check}
\acro{MIC}{Message Integrity Check}
\acro{PCM}{Pulse-code Modulation}
\acro{AHB}{Advanced High-performance Bus}
\acro{LSB}{Least Significant Bit}
\acro{TRM}{Technical Reference Manual}
\acro{IVT}{Interrupt Vector Table}
\acro{NVIC}{Nested Vectored Interrupt Controller}
\acro{NMI}{Non-maskable Interrupt}
\acro{FPB}{Flash Patch and Breakpoint}
\acro{SDIO}{Secure Digital Input Output}
\acro{ACL}{Asynchronous Connection-Less}
\acro{SCO}{Synchronous Connection-Orientated}
\acro{SP}{Stack Pointer}
\acro{FIFO}{First-In-First-Out}
\acro{CLI}{Command Line Interface}
\acro{API}{Application Programming Interface}
\acro{UAP}{Upper Address Part}
\acro{LAP}{Lower Address Part}
\acro{SDR}{Software-Defined Radio}
\acro{RF}{Radio Frequency}
\acro{DMA}{Direct Memory Access}
\acro{TID}{Transaction ID}
\acro{SSP}{Secure Simple Pairing}
\acro{TLS}{Transport Layer Security}
\acro{JW}{Just-Works}
\acro{NC}{Numeric-Comparison}
\acro{OOB}{Out-Of-Band}
\acro{MITM}{Man-In-The-Middle}
\acro{IP}{Internet Protocol}
\acro{PAN}{Personal Area Network}
\acro{ECDH}{Elliptic Curve Diffie-Hellman}
\acro{AES}{Advanced Encryption Standard}
\acro{CCM}{Counter mode with CBC-MAC}
\acro{ISM}{Industrial, Scientific and Medical}
\acro{LE SC}{LE Secure Connections}
\acro{PHY}{Physical Layer}
\acro{GATT}{Generic Attribute Profile}
\acro{MAC}{Media Access Control}
\acro{AFH}{Adaptive Frequency Hopping}
\acro{CVE}{Common Vulnerabilities and Exposure}
\acro{BPCS}{Broadcom Proprietary Control Signaling}
\acrodefplural{SDR}[SDRs]{Software-defined Radios}
\acrodefplural{LM}[LMs]{Link Managers}
\acrodefplural{CLI}[CLIs]{Command Line Interfaces}

\end{acronym}

\copyrightyear{2019}
\acmYear{2019}
\setcopyright{acmlicensed}
\acmConference[WiSec '19]{12th ACM Conference on Security and Privacy in Wireless and Mobile Networks}{May 15--17, 2019}{Miami, FL, USA}
\acmBooktitle{12th ACM Conference on Security and Privacy in Wireless and Mobile Networks (WiSec '19), May 15--17, 2019, Miami, FL, USA}
\acmPrice{15.00}
\acmDOI{10.1145/3317549.3319727}
\acmISBN{978-1-4503-6726-4/19/05}

\acmSubmissionID{95}

\usepackage[colorinlistoftodos,prependcaption,disable]{todonotes} 
\presetkeys%
    {todonotes}%
    {inline,backgroundcolor=orange}{}
\begin{document}

%
\title{Inside Job: Diagnosing Bluetooth Lower Layers Using Off-the-Shelf Devices}


\author{Jiska Classen}
\email{jclassen@seemoo.de}
\author{Matthias Hollick}
\email{mhollick@seemoo.de}
\affiliation{%
  \institution{TU Darmstadt, Secure Mobile Networking Lab}
 \city{Darmstadt}
 \country{Germany}
}

%
\renewcommand{\shortauthors}{Classen and Hollick} 

%
\begin{abstract}
Bluetooth is among the dominant standards for wireless short-range communication with multi-billion Bluetooth devices shipped each year. 
Basic Bluetooth analysis inside consumer hardware such as smartphones can be accomplished observing the \ac{HCI} between the operating system's driver and the Bluetooth chip.
However, the \ac{HCI} does not provide insights to tasks running inside a Bluetooth chip or \ac{LL} packets exchanged over the air. 
As of today, consumer hardware internal behavior can only be observed with external, and often expensive tools, that need to be present during initial device pairing. 
In this paper, we leverage standard smartphones for on-device Bluetooth analysis and reverse engineer a diagnostic protocol that resides inside Broadcom chips. Diagnostic features include sniffing lower layers such as \ac{LL} for Classic Bluetooth and \ac{BLE}, transmission and reception statistics, test mode, and memory peek and poke.
\end{abstract}

%
%

%

%
\maketitle


\section{Introduction}

Bluetooth is a widespread standard for short-range wireless communication. Initially developed as infrared replacement and for applications like wireless headphones in the '90s, it is now spreading with \ac{IoT} devices~\cite{bluetooth-history}.
Features still making it interesting nowadays are \ac{BLE} introduced in version 4.0, mesh networking in 5.0 and localization in 5.1~\cite{2016:SIG, 2017:SIG, 2019:SIG}.

A conventional Bluetooth diagnostic and sniffing setup requires external components and might necessitate wireless network infrastructure changes. A sniffer follows the frequency hopping scheme of a connection. Some devices only require basic connection establishment without pairing for encryption. To decrypt paired devices' encryption, a sniffer must observe the initial \ac{SSP} procedure, and if numeric comparison is used actively alter exchanged packets. Existing sniffing solutions range from professional but expensive Ellisys equipment to numerous  open source solutions, the latter having less stable implementations to follow encryption and hopping, such as Ubertooth and Bluefruit~\cite{ellisys-bv1, ubertooth_attify, bluefruit}. None of these run on the analyzed off-the-shelf device itself.

Most likely users of such a setup are aware of sniffing. We assume sniffers are installed to legally observe and analyze traffic between devices the analyst owns. Typical use cases are to inspect security of a proprietary smartphone app communicating with a proprietary \ac{IoT} device, or to analyze performance on the \ac{PHY} of a smartphone app and \ac{IoT} firmware under development.

Android devices offer the BTSnoop Log, a developer option, which only covers \ac{HCI} containing messages exchanged between the operating system and the Bluetooth chip. \ac{HCI} is the Bluetooth middleware layer, but lower layer packets are not directly encapsulated within \ac{HCI} and hence cannot be observed this way. In contrast, traces sniffed over the air with an external sniffer only contain Classic Bluetooth \ac{LMP} and \ac{BLE} \ac{LCP} packets. The toolchain implemented in this paper captures both, \ac{HCI} and \ac{LL} traces, and can inject \ac{LMP} packets. To the best of our knowledge, Bluetooth \acp{LL} did not experience security research yet, and we are the first to uncover major security issues within their implementations.

When reverse engineering symbols extracted from a Broadcom \ac{IoT} \ac{BLE}/\ac{BR} 5.0 evaluation kit firmware from 2018, we found an undocumented diagnostic serial protocol. A small subset of this diagnostic protocol is contained in Apple's Bluetooth Explorer and Packet Logger on some MacBooks~\cite{applebt}, but only allows some pre-configured usage for \ac{LL} sniffing and is not further documented. Diagnostic features also include memory reading and writing as well as dumps inside the firmware, producing statistics for throughput, and controlling device under test mode.

Broadcom is ranked market leader for wireless communication chips, followed by Qualcomm and MediaTek~\cite{gartnerwireless}. The diagnostic protocol that we reverse engineered is not disabled on consumer devices and can be found throughout all smartphones with Broadcom chips we analyzed, such as Nexus 5/6P and Samsung Galaxy S6. Moreover it is supported by chips in Raspberry Pi 3/3+.

\begin{figure*}[htp]
	\begin{center}
	\begin{tikzpicture}[minimum height=0.5cm, scale=0.8, every node/.style={scale=0.8}, node distance=0.55cm]
		\node[draw,  minimum width=4cm] (rfcomm) {RFCOMM};
		\node[draw, right=of rfcomm, minimum width=3.3cm] (sdp) {SDP};
		\node[draw, below=of rfcomm.west, minimum width=8cm, anchor=west] (l2cap) {L2CAP};
		\node[draw, below=of l2cap.west, minimum width=8cm, minimum height=1cm, anchor=west, 
				yshift=-0.3cm, align=center] (hci) {Host Controller Interface (HCI)\\via UART};
		\node[draw, below=of hci.west, minimum width=4cm, anchor=west, yshift=-0.3cm] (devmgr) {Device Manager};
		\node[draw, right=of devmgr, minimum width=3.3cm] (linkmgr) {Link Manager};
		\node[draw, below=of devmgr.west, minimum width=8cm, anchor=west] (brm) {Baseband Resource Manager};
		\node[draw, below=of brm.west, minimum width=8cm, anchor=west] (linkctrl) {Link Controller};
		\node[draw, below=of linkctrl.west, minimum width=8cm, anchor=west] (phy) {Bluetooth PHY};
		\draw[thick] (rfcomm.south) -- (rfcomm.south |- l2cap.north);
		\draw[thick] (sdp.south) -- (sdp.south |- l2cap.north);
		\draw[thick] (l2cap.south) -- (hci.north);
		\draw[thick] (devmgr.north |- hci.south) -- (devmgr.north);
		\draw[thick] ([xshift=0.5cm] hci.south) -- ([xshift=0.5cm] brm.north);
		\draw[thick] (linkmgr.north |- hci.south) -- (linkmgr.north);
		\draw[thick] (devmgr.south) -- (devmgr.south |- brm.north);
		\draw[thick] (linkmgr.south) -- (linkmgr.south |- brm.north);
		\draw[thick] (brm.south) -- (linkctrl.north);
		\draw[thick] (linkctrl.south) -- (phy.north);

		\draw[dashed] ([xshift=-3cm] hci.west) -- ([xshift=0.5cm] hci.east);
		\node[left=of hci.west, anchor=west, xshift=-2.2cm, yshift=1.7cm] (host) {Host};
		\node[left=of hci.west, anchor=west, xshift=-2.2cm, yshift=-0.5cm]  (controller) {Controller};
		\node[left=of hci.west, anchor=west, xshift=12.4cm, yshift=1.5cm, align=left] (remote) {Remote\\Device};
		\node[left=of hci.west, anchor=west, xshift=-8.3cm, yshift=1.5cm, align=left] (diagctl) {Diagnostics\\Control};
		
		\node[inner sep=0pt] (hostimg) at (-3.3,-0.7)
    {\includegraphics[height=1.7cm]{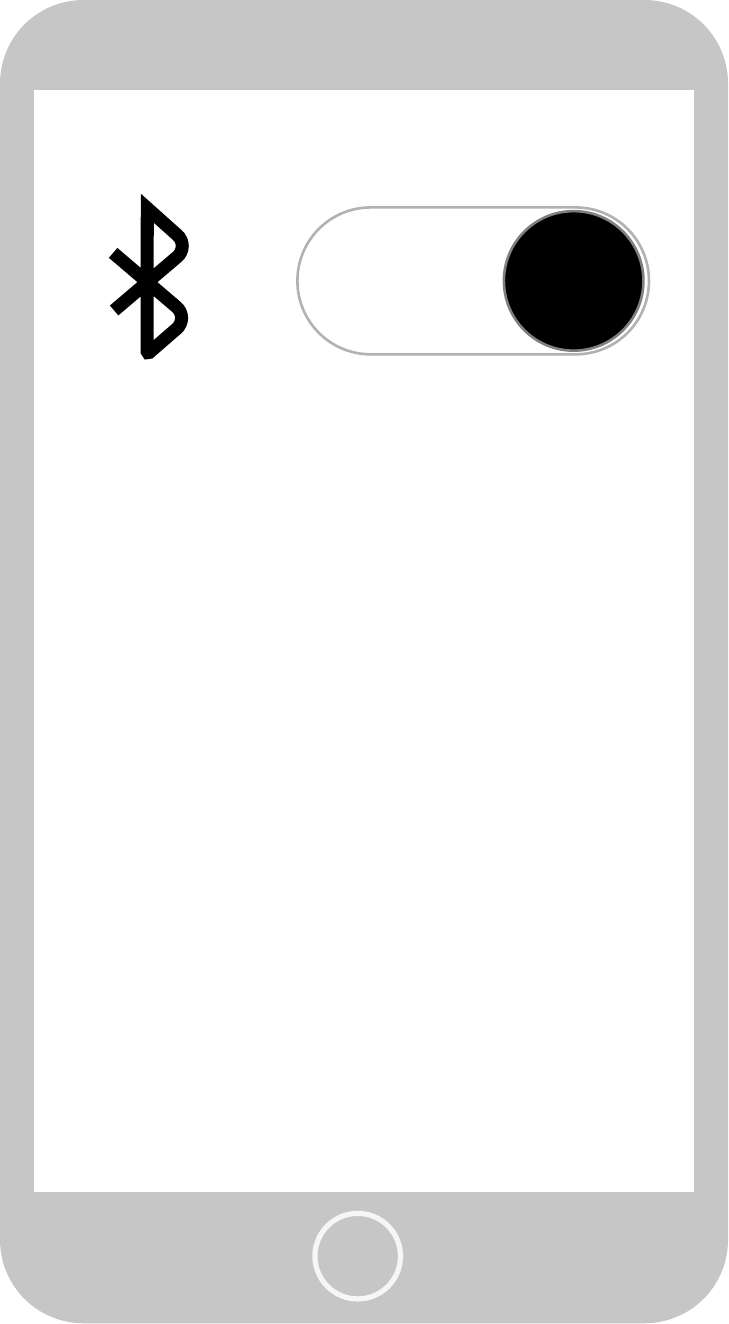}};
		\node[inner sep=0pt] (remoteimg) at (9,-0.7)
    {\includegraphics[height=1.7cm]{pics/host.pdf}};
		\node[inner sep=0pt] (controllerimg) at (-3.3,-3.3)
    {\includegraphics[width=2cm]{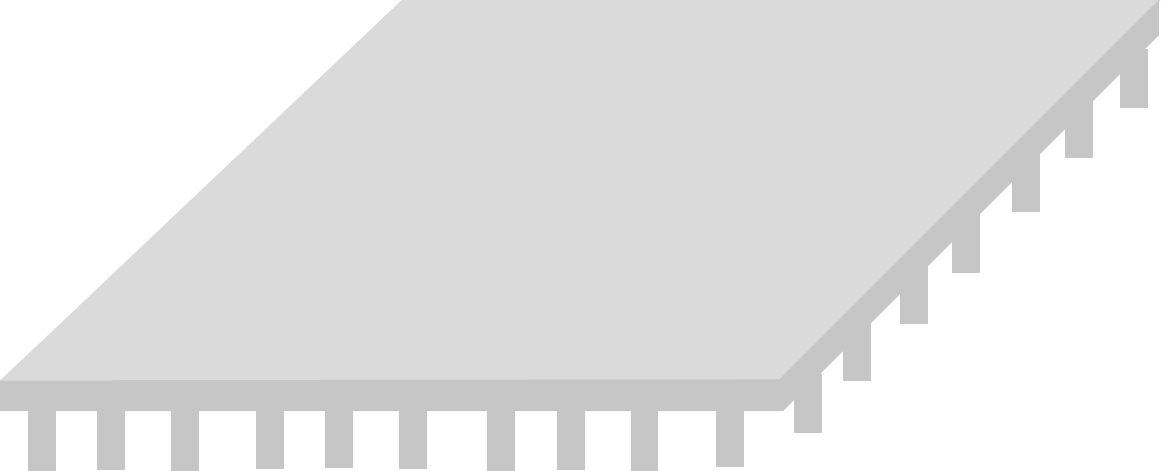}};
		\node[inner sep=0pt] (laptopimg) at (-8.4,-0.7)
    {\includegraphics[height=1.7cm]{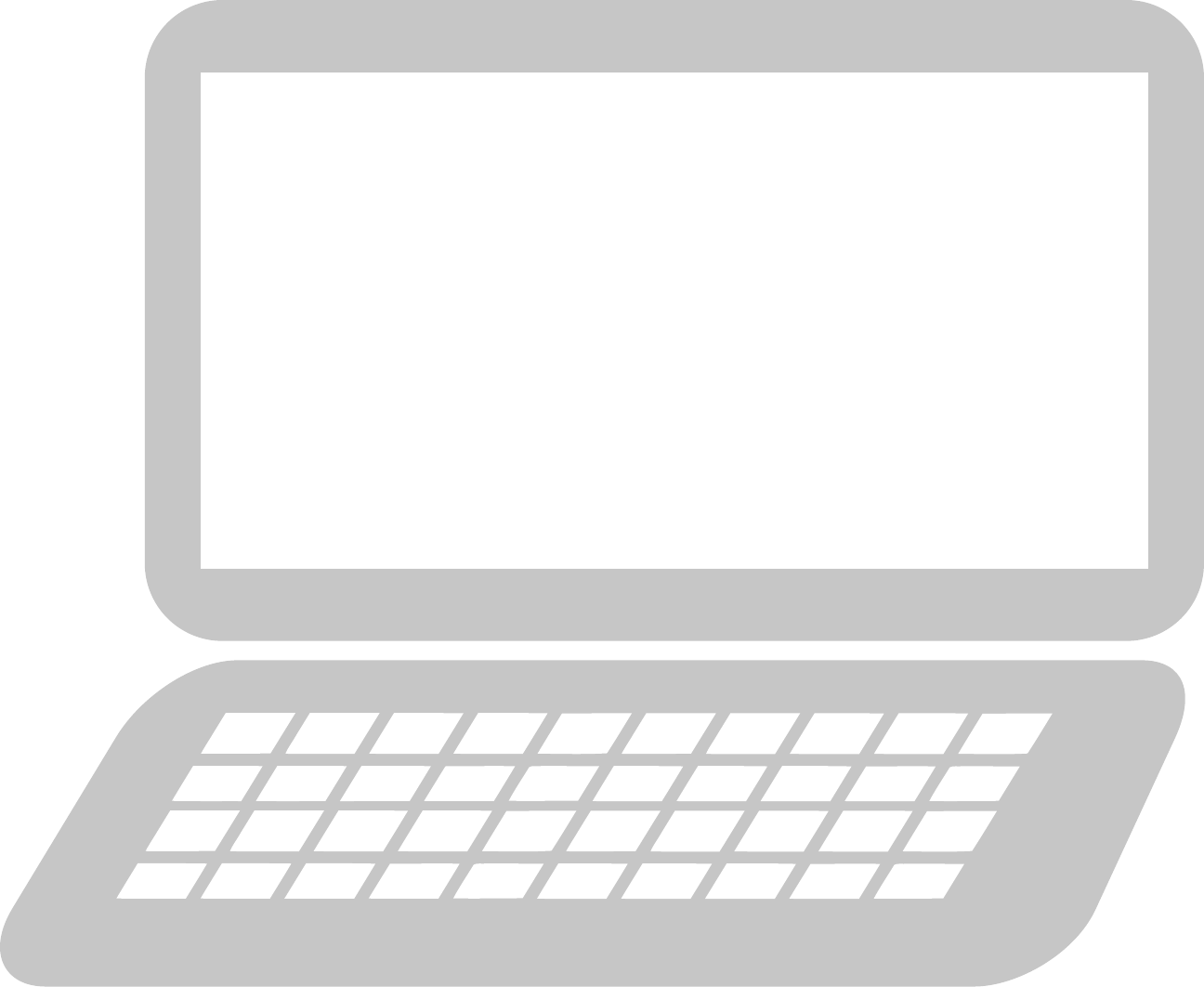}};
    
		\draw[dotted] (phy.east) -- (remoteimg.south |- phy.east) -- (remoteimg.south);
		\draw[dotted] (hostimg.west) -- (laptopimg.east);
		\node[left=of laptopimg.east, anchor=west, align=left, xshift=0.8cm, yshift=-0.2cm] (adb) {ADB};
	\end{tikzpicture}
	\end{center}
	\vspace{-1em}
	\caption{Architecture of the Bluetooth protocol stack including diagnostics setup.}
	\label{fig:bluetooth_architecture}
\end{figure*}

With this paper, we enable diagnostic mode on a variety of off-the-shelf devices and program software to control and interpret diagnostics. Our contributions are as follows:
\begin{itemize}
\item Implementation of live forwarding of diagnostic messages to a Linux host which can remotely control and analyze diagnostic mode including a live traffic view.
\item Patching drivers of Android 6.0.1, Android 7.1.2 and LineageOS 14.1 to forward the diagnostic protocol. 
\item Enabling \ac{LMP} injection.
\item Showcase this is a powerful security analysis tool by discovering two \acp{CVE} based on malicious \ac{LMP} packets.
\end{itemize}

Source code for using Broadcom diagnostics with \emph{InternalBlue} is publicly available on \url{https://github.com/seemoo-lab/internalblue}. It is the first zero cost sniffer for the Bluetooth \ac{LL}.

This paper is structured as follows. \autoref{sec:background} gives an overview on the Bluetooth protocol stack and explains how to utilize it. In \autoref{sec:reversing}, Broadcom's diagnostic protocol features and inner workings are reversed. We apply this knowledge on the Android device, controllable and observable with Python scripts and Wireshark on a Linux host connected over \ac{ADB}, and make first experiments in diagnostic mode including security flaw detection in \autoref{sec:implementation}. We discuss the outcome and Broadcom's protocol design in \autoref{sec:discussion}. We conclude our findings in \autoref{sec:conclusion}.

%
%

\section{System Overview}
\label{sec:background}

\autoref{fig:bluetooth_architecture} gives an overview of the Bluetooth protocol stack and hardware components involved. The host system, in our case Android or LineageOS, has a Bluetooth driver providing communication services to applications. The controller, in our case a Broadcom chip, manages connections and performs wireless communication.

\begin{figure}[hbp]
\centering
\vspace{-1em}
  \includegraphics[width=\columnwidth]{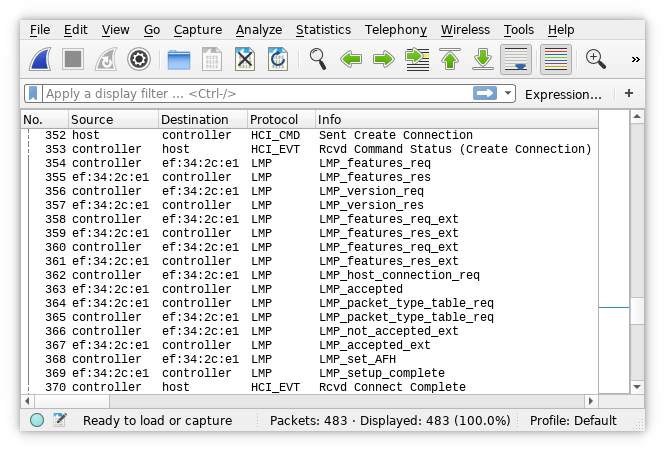}
  \vspace{-1.5em} 
  \caption{Connection establishment trace with \ac{HCI} and \ac{LMP} between two Nexus 5 smartphones.}
  \label{fig:lmp_android}
\end{figure}

A diagnostic trace including \ac{HCI} and \ac{LMP} packets can be found in \autoref{fig:lmp_android}. It is produced on Android by activating \ac{LL} diagnostics through our custom patched Bluetooth driver and interpreting the output with our Wireshark dissector. Even though Bluetooth \ac{MAC} addresses have the standard 6 byte format, only the lower 4 bytes are required to address devices~\cite{2009:ossmann}. This is why diagnostic traces do not include the full \ac{MAC} address.

\subsection{Interface Between Operating System and Bluetooth Chip}
Host and controller are connected via \ac{HCI} as defined in the Bluetooth standard~\cite[p.~656]{2016:SIG}. The host sends \ac{HCI} commands to the controller, while the controller indicates \ac{HCI} events happening. Events can be replies to commands, but they can also happen asynchronous. 
In \autoref{fig:lmp_android}, the host initiates a connection and the controller confirms this directly with an event. Later this will be answered with a connect complete event, which can confirm successful connection establishment but also indicate a page timeout. The host is not aware of any \ac{LL} actions performed between requesting connection establishment and the feedback if this succeeded.

\ac{HCI} is controllable via a serial interface using \ac{UART}. On Android systems it appears as \path{/dev/HS99} or a similar number. For testing purposes the Android driver can be bypassed by directly writing to \ac{UART}. A useful \ac{HCI} command to determine if the controller is running on a Broadcom chip is reading local version information, which can be sent on an Android shell as follows:
\begin{lstlisting}
echo -ne '\x01\x01\x10\x00' >/dev/ttyHS99
\end{lstlisting}
The first byte is set to \texttt{0x01} to indicate in an H4 header that the message will be an \ac{HCI} command, as listed in \autoref{tab:h4types}. Following content is a standard compliant read local version information command. The controller answers with a command complete event containing version information.

\begin{table}[b]
	\small
	\begin{tabular*}{\linewidth}{ @{\extracolsep{\fill}} p{4cm} r l}
	
		\textbf{H4 Type} & \textbf{Command} & \textbf{Direction} \\
		\hline		
		HCI Command & \texttt{0x01} & Host $\rightarrow$ Controller \\
		ACL Data & \texttt{0x02} & Host $\leftrightarrow$ Controller \\
		SCO Data & \texttt{0x03} & Host $\leftrightarrow$ Controller \\
		HCI Event & \texttt{0x04} & Host $\leftarrow$ Controller \\
		\hline
		Broadcom Diagnostics & \texttt{0x07} &  Host $\leftrightarrow$ Controller\\
		Message Queue Put & \texttt{0x0a} & Host $\leftrightarrow$ Controller \\
		WICED Studio (evaluation kit only) & \texttt{0x19} &  Host $\leftrightarrow$ Controller \\
	\end{tabular*}
	\caption{H4 field bytes indicating \ac{UART} traffic types.}
	\label{tab:h4types}
\end{table}

The Android driver offers more convenient ways to control this serial interface. When compiled with the flag \lstinline{BT_NET_DEBUG=TRUE}, it locally opens \ac{TCP} port 8872 for sniffing and 8873 for injecting messages~\cite{aospnetwork}. Traffic to these ports can be forwarded over \ac{ADB} to control them from a remote device. Message contents are restricted to H4 types \texttt{0x01}-\texttt{0x04} without Android source code changes.

\subsection{Over-the-Air Communication Between Bluetooth Devices}

Bluetooth devices use a \ac{PHY} which is frequency hopping on 79 channels. The Bluetooth master in a piconet controls the hopping clock for all slaves. Communication can be encrypted based on keys exchanged with \ac{ECDH}. Details of lower layers are abstracted from the host and only handled inside the controller. An overview of connection establishment steps followed by diagrams of \ac{HCI} and \ac{LMP} packets exchanged can be found in the Bluetooth specification~\cite[p.~1420]{2016:SIG}.

Wireless Bluetooth sniffers aim to extract management protocols exchanged over the air. On Classic Bluetooth this protocol is \ac{LMP}, while it is \ac{LCP} for \ac{BLE}. To successfully decrypt wireless message contents, an over-the-air sniffer must be present during initial pairing and succeed in following the hopping sequence. If the latter fails, \ac{SSP} must be manually repeated multiple times.
\todo{if space allows: explain a bit more about ssp}

A sniffer implemented inside the Bluetooth controller is not required to observe the initial pairing for demodulation and decryption. The controller knows all necessary connection information and keeps internal states of not yet modulated, respectively demodulated, \ac{LMP} and \ac{LCP} messages. This is the case for the setup in this paper. Observing messages inside the Bluetooth controller comes with the challenge to locate and extract this information.

\section{Reversing the Diagnostic Protocol}
\label{sec:reversing}

Broadcom enables testing their newest chipsets with evaluation boards. The most recent available Bluetooth platform is \emph{CYW920735 Q60EVB-01}, 
an \ac{IoT} BLE/BR 5.0 evaluation kit~\footnote{Cypress recently acquired some \ac{IoT} branches of Broadcom, but despite the name change firmware and features are similar.}, which is not only meant to offer Bluetooth connectivity but also has sufficient memory to run an \ac{IoT} application within the same chip. This application can be programmed in WICED Studio~\cite{wiced}.
When reversing firmware of the evaluation kit we realized that the custom application thread for the \ac{IoT} platform is using the same thread setup functions as existing threads within the Bluetooth firmware, suggesting that there might be common bindings for several functions inside the Bluetooth firmware within WICED Studio. In fact, a file called \path{patch.elf} contains all symbols of the evaluation kit Bluetooth chip's firmware \emph{CYW20735B01}. In addition the library \path{wiced_hidd_lib.a} contains struct and parameter names, but correct mappings are harder to locate in this format.

\todo{Müssen/sollten wir etwas sagen, das das legal ist unter unserer Rechtsprechung? Ggf. auch hier nochmal sagen, das für fortgeschrittenen Sicherheitsanalyse es notwendig sein kann/wichtig ist XYZ zu beobachten und wir aus diesem Grund eine analyse des codes via reverse engineering vorgenommen haben.}

\subsection{Accessing the Link Layer}
Being interested in lower layers, the function names \lstinline{diag_log}\lstinline{TxLmpPkt},
\lstinline{diag_logRxLmpPkt} and \lstinline{diag_logLcpPkt} stand out.
All of these first check if \lstinline{diag_sendLmpPktFlag} is set,
and then copy data to call the function \lstinline{tran_sendDiagDataToHost}. However, the opposite data flow must be found to enable logging. When checking references to the flag enabling sending \ac{LL} diagnostic logging data to the host, the function \lstinline{diag_processCommand} appears. The diagnostic command \texttt{0xf0} followed by \texttt{0x01} enables logging, while \texttt{0xf0} followed by \texttt{0x00} disables logging.

Further analysis of \lstinline{diag_processCommand} shows that enabling \ac{LL} logging is not the only diagnostic command. An overview of all commands and answers generated is shown in \autoref{tab:bcmdiag}. An incomplete list of commands and their meanings is contained in Apple's Packet Logger, which is part of the additional tools for Xcode. Not all MacBooks support \ac{LL} logging with Xcode tools. In our experiments it worked on a MacBook Pro from 2011, but could not be enabled on a MacBook Pro from 2016.

\begin{table}[b]
	\small
	\begin{tabular*}{\linewidth}{ @{\extracolsep{\fill}} l r l}
	
		\textbf{Diagnostic Feature} & \textbf{Command} & \textbf{Direction} \\
		\hline		
		  LMP Sent & \texttt{0x00} & Host $\leftarrow$ Controller \\
		  LMP Received & \texttt{0x01} & Host $\leftarrow$ Controller \\
		  Memory Access Response to Peek & \texttt{0x03} & Host $\leftarrow$ Controller \\
		  Memory Hex Dump Response & \texttt{0x04} & Host $\leftarrow$ Controller \\
		  Reported Completed Test & \texttt{0x0a} & Host $\leftarrow$ Controller \\
		  Memory Access Response to Poke& \texttt{0x11} & Host $\leftarrow$ Controller \\
		  CPU Load Response & \texttt{0x15} & Host $\leftarrow$ Controller \\
		  Basic Rate ACL Stats Data & \texttt{0x16} & Host $\leftarrow$ Controller \\
		  EDR ACL Stats Data & \texttt{0x17} & Host $\leftarrow$ Controller \\
		  Received AUX Response & \texttt{0x18} & Host $\leftarrow$ Controller \\
		  SCO Stats Data (Type \texttt{0x1a}) & \texttt{0x1a} & Host $\leftarrow$ Controller \\
		  eSCO Stats Data (Type \texttt{0x1b}) & \texttt{0x1b} & Host $\leftarrow$ Controller \\
		  Get Connection Response & \texttt{0x1f} & Host $\leftarrow$ Controller \\
		  LCP Sent & \texttt{0x80} & Host $\leftarrow$ Controller \\
		  LCP Received & \texttt{0x81} & Host $\leftarrow$ Controller \\
		  Reset Basic Rate ACL Stats & \texttt{0xb9} & Host $\rightarrow$ Controller \\
		  Get Basic Rate ACL Stats & \texttt{0xc1} & Host $\rightarrow$ Controller \\
		  Get EDR ACL Stats & \texttt{0xc2} & Host $\rightarrow$ Controller \\
		  Get AUX Stats & \texttt{0xc3} & Host $\rightarrow$ Controller \\
		  Get SCO Stats (Type \texttt{0x1a}) & \texttt{0xc5} & Host $\rightarrow$ Controller \\
		  Get eSCO Stats (Type \texttt{0x1b}) & \texttt{0xc6} & Host $\rightarrow$ Controller \\
		  Get Connection Stats & \texttt{0xcf} & Host $\rightarrow$ Controller \\
		  Toggle LMP Logging & \texttt{0xf0} & Host $\rightarrow$ Controller \\
		  Memory Peek & \texttt{0xf1} & Host $\rightarrow$ Controller \\
		  Memory Poke & \texttt{0xf2} & Host $\rightarrow$ Controller \\
		  Memory Hex Dump & \texttt{0xf3} & Host $\rightarrow$ Controller \\
		  Run Test & \texttt{0xf6} & Host $\rightarrow$ Controller \\
	\end{tabular*}
	\caption{Complete list of diagnostic features and their command codes on \emph{CYW20735B01}.}
	\label{tab:bcmdiag}
\end{table}

\begin{figure*}[htp]
\captionsetup[subfigure]{justification=centering}
\centering
\begin{subfigure}{.42\textwidth}
  \centering
  \includegraphics[height=8.3cm]{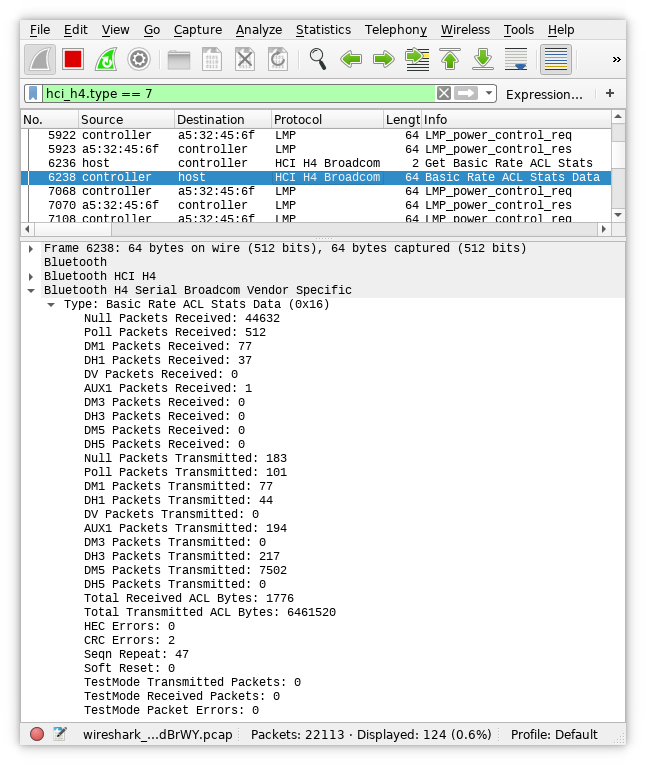}
\vspace{-0.5em}
  \caption{Live \ac{BR} \ac{ACL} \\statistics while a connection to a headset is running.}
  \label{fig:wireshark-bracl}
\end{subfigure}\hfill%
\begin{subfigure}{.55\textwidth}
  \centering
  \includegraphics[height=8.3cm]{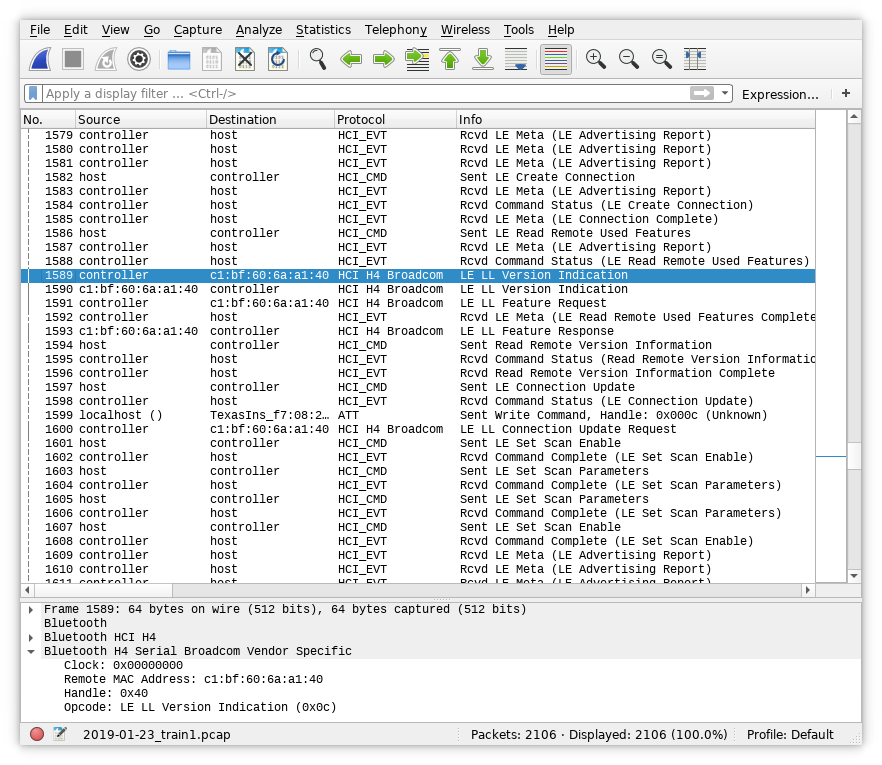} 
\vspace{-0.5em}
  \caption{Recorded \ac{LCP} trace within a connection to a \\smart toy train equipped with \ac{BLE}.}
  \label{fig:wireshark-trainle}
\end{subfigure}
\vspace{-0.5em}
\caption{Wireshark Broadcom diagnostics example traces recorded on a Nexus 5.}
\label{fig:test}
\end{figure*}

\subsection{Diagnostic Command Processing}
\label{ssec:diagcmd}
The function actually receiving diagnostic commands turns out to be \lstinline{btuarth4_HandleRX}\lstinline{FullMsgDone}, which does not only parse H4 types \texttt{0x01}-\texttt{0x04}, but accepts one documented message type \texttt{0x19} for WICED application debugging and the undocumented diagnostic message type \texttt{0x07}. Upon receiving a message starting with \texttt{0x07}, the function \lstinline{btuarth4_process_diag} is called, which then calls the function of interest, \lstinline{diag_processCommand}. A further undocumented H4 communication channel for putting messages into send queues is \texttt{0x0a}, but we could only find it on Raspberry Pi 3/3+. \autoref{tab:h4types} also lists Broadcom proprietary H4 types.

\subsubsection*{Diagnostic Control Commands}
The biggest challenge is that some diagnostic commands and message contents are unknown and unnamed. The Wireshark dissector developed in \autoref{ssec:implwireshark} can be optimized by programming a first version, recording traces, analyzing missing fields, adding them to the dissector, and then recording another round of traces. Debugging and reversing of further fields takes a lot of time and some of their contents are still to be guessed, such as the statistic types \texttt{0x1a} and \texttt{0x1b} in \autoref{tab:bcmdiag}. Nonetheless, we support decoding most of the outputs including \ac{BR} and \ac{EDR} statistics with an example given in \autoref{fig:wireshark-bracl}. Statistics summarize packets transmitted and received along with errors occurred---now observable even without continuous monitoring.

\subsubsection*{Memory Access}
Diagnostic memory access is similar to what is provided over \ac{HCI}.
Documented \ac{HCI} commands are \lstinline{Read_RAM} and \lstinline{Write_RAM}~\cite{cypress_cyw43438}, which can modify memory of the ARM core inside the chip.
An undocumented \ac{HCI} command that can modify the Bluetooth modem BlueRF memory is \lstinline{SuperDuperPeekPoke} (\texttt{0xfc0c}). The Bluetooth modem is a separate component on the chip, which is responsible for modulation.

Diagnostic memory peek (\texttt{0xf1}) and memory poke (\texttt{0xf2}) allow reading and writing single memory bytes. The first memory peek and poke argument is the access type, which can be ARM memory (\texttt{0x02}) or BlueRF modem memory (\texttt{0x03}). The second argument is a 4 byte reverse byte order address. For memory poke, this is followed with the 1 byte value written to this address.
Memory hex dump (\texttt{0xf3}) needs always to be passed access type ARM memory (\texttt{0x04}) and a 4 byte address. It returns 32 bytes memory content starting from the address passed.

Diagnostic memory access has fewer message overhead and parsing steps and therefore suits better than \ac{HCI} for observing variable contents while debugging firmware behavior.

\subsubsection*{Test Mode}

The test command (\texttt{0xf6}) configures a test with similar parameters as \lstinline{LMP_test_control} over the air. More interesting is the possibility to get a summary of a test status. The test status is received within the diagnostic test response (\texttt{0x0a}) and includes the number of transmitted and received test packets. To get this basic test information no external test equipment is required.
\todo{test the above ... should be 1 byte status, 2 bytes test tx, 2 bytes zero, 2 bytes test rx}

\todo{params to execute the test itself are not correct so far :( todo: set tracepoints at the function that receives the connection pointer and try to get correct params... crash is too early for tp within the actual error}

\subsection{Compatibility}
After searching for similar H4 message type switch cases in other Broadcom Bluetooth firmware versions it turns out that \lstinline{btuarth4_HandleRX} is implemented throughout multiple firmwares, such as \emph{BCM2070B0} (MacBook Pro 15'' early 2011, Lenovo Thinkpad T420), \emph{BCM4335C0} (Nexus 5, Xperia Z3 Compact, Samsung Galaxy Note 3), \emph{BCM43430A1} (Raspberry Pi 3), \emph{BCM4345C0} (Raspberry Pi 3+), \emph{BCM4358A3} (Nexus 6P, Samsung Galaxy S6, Samsung Galaxy S6 edge), and \emph{CYW20735B01} (\ac{IoT} BLE/BR 5.0 evaluation kit). Their build dates range from 2008 to 2018. We do not have a complete set of Broadcom and Cypress Bluetooth firmwares but assume this feature is present on all of their chips.

\section{Implementation}
\label{sec:implementation}

Reversing diagnostic commands is only the first step in using them on off-the-shelf devices. Further required steps are patching the operating system's driver to forward and process diagnostic messages, as explained in \autoref{ssec:implandroid}, displaying them human readable in a Wireshark dissector, described in \autoref{ssec:implwireshark}, using them in active Bluetooth connections to do first tests (\autoref{ssec:measure}), and based on intermediate results improve the Wireshark dissector. Finally, we uncover security issues within \ac{LMP} parsing inside Broadcom firmware we detected on various chips in \autoref{ssec:injectsecurity}.

\subsection{Patching Bluetooth on Android}
\label{ssec:implandroid}

On Android 6.0.1, Android 7.2.1 and LineageOS 14.1, the Bluetooth stack is implemented similarly. The biggest design problem is that H4 types from \autoref{tab:h4types} as well as their directions and lengths are hardcoded in multiple places. This makes the required changes 44 lines of code in Android 6.0.1, and 41 lines of code in Android 7.2.1 and LineageOS 14.1. Again, the driver needs to be compiled with the flag \lstinline{BT_NET_DEBUG=TRUE} to enable injection and sniffing as described in \autoref{sec:background}. A properly compiled Bluetooth driver results in a file \path{bluetooth.default.so}, which can be copied to the according system folder as privileged user. 

Android 8 introduced major changes to the Bluetooth driver. The basic driver is still using similar code. In addition, interfaces are introduced---for components that should not be changed, since changes are \ac{ABI} breaking and might stop other programs from working properly. The code of the new interfaces \lstinline{IBluetoothHci} and \lstinline{IBluetoothHciCallbacks} defines functions and callbacks for H4 types \texttt{0x01}-\texttt{0x04}~\cite{aospbtinterface}. This means introducing new H4 types to Android 8 is an \ac{ABI} breaking change, so we decided to currently not support it.

\subsection{Wireshark Dissector}
\label{ssec:implwireshark}
Programming a Wireshark dissector for the Broadcom diagnostic protocol creates some challenges. To ensure a good performance along with portability, we implement the dissector as a plugin in C.

As mentioned in \autoref{ssec:diagcmd}, the actual challenge in writing a dissector is to reverse engineer undocumented fields. We were able to reverse engineer almost all fields of the diagnostic protocol, the only missing part being the meanings of \ac{SCO} and \ac{AUX} statistics.

Luckily the \acp{LL}, \ac{LMP} for Classic Bluetooth and \ac{LCP} for \ac{BLE}, are described in the Bluetooth specification~\cite[p.~508,\,p.~2589]{2016:SIG}.

For \ac{LMP}, a C-based dissector that can be integrated already exists~\cite{btbrlmp}. Unfortunately the Broadcom diagnostic protocol deviates from specification: it always reports the maximum length of 17 bytes for received packets. This needs to be fixed before passing these packets to the existing \ac{LMP} dissector. Moreover, Broadcom implements an undefined \ac{LMP} packet type \texttt{0x00} which is used for \ac{BPCS}, as listed in \autoref{tab:bpcs}. This feature is only used within Broadcom Bluetooth connections, we were able to observe it during a Nexus 5 and Samsung Galaxy S8 pairing. A sample \ac{LMP} trace was already explained for \autoref{fig:lmp_android}.

Dissectors can also be found for \ac{LCP}, but they are either not C-based or not accepting the same header format, so we decided to implement a dissector from scratch. Broadcom again added their own features, this time using \ac{LCP} type \texttt{0xff} with meanings listed in \autoref{tab:leext}. A trace for a \ac{BLE} enabled child toy train that can make sounds and change driving speed with an app is shown in \autoref{fig:wireshark-trainle}. Overall, \ac{LCP} is more simplistic than \ac{LMP} and requires fewer messages exchanged to establish and manage connections. Traces include the full \ac{MAC} address of the connected device.

\begin{table}[b]
	\small
	\begin{tabular*}{\linewidth}{ @{\extracolsep{\fill}} p{2.5cm} p{6cm}}
	
		\textbf{LMP BPCS Type} & \textbf{Purpose} \\
		\hline
		\texttt{0x00}	& Features Request \\
		\texttt{0x01}	& Features Response \\
		\texttt{0x02}	& Not Accept \\
		\texttt{0x03}	& BFC Suspend \\
		\texttt{0x04}	& BFC Resume Request/Response \\
		\texttt{0x05}	& BPCS Accept \\
		\hline
		\texttt{0x95}	& Enable Device Under Test Mode (Vulnerability) \\
	\end{tabular*}
	\caption{\ac{LMP} vendor specific BPCS packet types.}
	\label{tab:bpcs} 
	\vspace{-1.5em} 
\end{table}

\begin{table}[b]
	\small
	\begin{tabular*}{\linewidth}{ @{\extracolsep{\fill}} p{2.5cm} p{6cm}}
	
		\textbf{LCP Extended Type} & \textbf{Purpose} \\
		\hline
		\texttt{0x01}	& Vendor Specific Feature Request \\
		\texttt{0x02}	& Vendor Specific Feature Response \\
		\texttt{0x03}	& Vendor Specific Enable BCS Timeline \\
		\texttt{0x04}	& Random Address Change \\
	\end{tabular*}
	\caption{\ac{LCP} vendor specific packet types.}
	\vspace{-1.5em}
	\label{tab:leext}
\end{table}

\subsection{Performance}
\label{ssec:measure}

While enabling diagnostic logging or making statistic and memory requests we did not recognize any performance issues or packet loss while a connected headset was playing music. Upon initialization of the \ac{UART} interface, the operating system updates the baud rate to 3\,Mbit/s. This result is not surprising as the maximum throughput of \ac{EDR} is also 3\,Mbit/s.

\todo{some sample statistics and other recordings ... ie make cross checks on packet numbers}

\todo{test mode might be better to put some nice numbers here}

\subsection{LMP Injection and Security Analysis}
\label{ssec:injectsecurity}

Broadcom not only implements an undocumented diagnostic H4 interface. The standard compliant way of allowing modifications are vendor specific \ac{HCI} commands---we were able to reconstruct 259 command names and some of their parameters. In the context of \ac{LL}, \lstinline{SendLmpPdu} (\texttt{0xfc58}) is the most interesting---it turns our diagnostic sniffer into an \ac{LMP} injector. \ac{LMP} packets are still validated against standard compliant opcodes and lengths when sent with this function, but except from this limitation any active connection handle can be selected as receiver. While testing this feature we already discovered two major security issues, CVE-2018-19860 and CVE-2019-6994, and reported these to Broadcom.

\subsubsection*{Security Issues within Broadcom Firmware}
The first security issue CVE-2018-19860  is related to boundary checking of \ac{LMP} opcodes. Handler tables define which function to execute upon receiving a packet with a certain opcode. When implementing the vendor specific \ac{BPCS} packet type, Broadcom forgot to apply an opcode range check. Any opcode larger than \texttt{0x05}, the maximum defined in \autoref{tab:bpcs}, potentially executes a function. Compilers tend to put multiple handler tables after each other. On Nexus 5 (\emph{BCM4335C0}), the \ac{LMP} \ac{BPCS} handler table is followed by \ac{HCI} handler tables. With this vulnerability, an attacker can issue \ac{HCI} commands over the air, which should only be possible as privileged user on the host. For example the \ac{LMP} \ac{BPCS} opcode \texttt{0x95} is mapped to the command \lstinline{HCI_enable_device_under_test_mode}. Executing test mode can be used make the target device jam selected frequencies and drains battery. Depending on the firmware version memory contents differ, but limited code execution is probably possible on most chips. If invalid addresses are executed, the Bluetooth of the device under attack simply crashes. Vulnerable devices include MacBook Pro 2016 and Raspberry Pi 3. Even the iPhone 6 was vulnerable to this issue until it was fixed in iOS 12.1.3.

The second vulnerability CVE-2019-6994 happens on out of order execution of \ac{LMP} commands. An attacker initiates \ac{SSP} but does not wait for user confirmation or cancellation of the pairing procedure. Instead he directly sends an
\lstinline{LMP_start_encryption_req}, which causes the
Bluetooth firmware of the device under attack to crash. The crash happens within the \lstinline{bignum_xormod}
calculation. This issue only is present on very old chips such as Nexus 5 (\emph{BCM4335C0}).

To exploit any of the vulnerabilities above, Bluetooth must be turned on. The attack can be
performed by anyone connecting to the device, previous pairing is not
required. An attacker only needs to know the MAC address, the device
can be invisible.

\subsubsection*{Commercial Bluetooth Stack Issues}
Discovery of the \ac{LMP} issues inspired us to implement an \ac{MAC} address based firewall that only allows connections from previously paired devices such as the user's headset. Any \ac{LMP} request originating from an untrusted \ac{MAC} address is answered with \lstinline{LMP_not_accepted}. This patch performs well on the protected device.

When testing this filter in the wild we found that on iOS devices Bluetooth restarts exactly one minute after trying to establish a connection to a firewalled device. The cause for this is bluetoothd, the Bluetooth host implementation on iOS, which decides to restart upon unexpected behavior---in this case a timeout from the \ac{HCI} create connection command, which does not return an \ac{HCI} event as expected. Restarting Bluetooth is not necessary in this context and terminates currently established connections. Most likely Apple implemented this as a workaround for instabilities inside the Bluetooth firmware. We reported this bug to Apple in December 2018 but did not get any feedback yet.

\section{Discussion}
\label{sec:discussion}

Integrating diagnostic features into a Bluetooth chip is something probably all Bluetooth chip manufacturers require in their development process.
Leaving these in production level firmware builds in combination with obscuring them by not documenting the corresponding interfaces can be considered a bad practice and only provides security by obscurity.

Our previous work already enables \ac{LMP} monitoring and injection on the \emph{BCM4335C0} chip in Nexus 5~\cite{dennisthesis}. We used assembly firmware patches and recompiled Android drivers, with the former making it hard to port to other devices. It did not include \ac{LCP} logging and the other diagnostic features listed in \autoref{tab:bcmdiag}.

Security issues in the \ac{LMP} handler were discovered while actually trying to reverse engineer and explore features. Code execution within \ac{LMP} \ac{BPCS} parsing was found when trying to figure out what this proprietary handler is doing. An \lstinline{LMP_start_encryption_req} was sent accidentally when confusing the maximum valid \ac{LMP} length field with the opcode field, which both are 17. Most likely Bluetooth \acp{LL} did not experience proper testing due to lack of openly available tools---a gap we aim to close with this paper.

\section{Conclusion}
\label{sec:conclusion}

Once reverse engineered, the Broadcom diagnostic protocol offers a lot of options and is easy to use. In future releases of Bluetooth stacks this could become a debug feature integrated by default.

Future work will be porting Broadcom diagnostics to further protocol stacks, such as BlueZ in Linux~\cite{bluez}. Linux integration should be straightforward, sockets for interception and injection are provided by default. Diagnostic commands are filtered, hence some code changes are required. Our reverse engineering results already confirm Broadcom diagnostics exist in Bluetooth chips of Raspberry Pi 3/3+, which are even cheaper than Android smartphones.

Bluetooth plays an important role in security critical application areas such as the \ac{IoT}. By reverse engineering the Broadcom diagnostic protocol and providing an open source analysis framework, we enable advanced on-device analysis of lower layer Bluetooth protocols---the remote device under test can be from any vendor.

During the development process, we already discovered critical vulnerabilities in the implementation of the \ac{LL} protocols in widely deployed Bluetooth firmware.
It is to be expected that \ac{LL} parsing in Bluetooth firmwares of other vendors has similar bugs. 
We encourage researchers to dig deeper into wireless firmware in general, especially if little is known or documented.


%

\begin{acks}

We thank Broadcom for providing patches to vendors.
This work has been funded by
the \textsc{dfg}
within \textsc{sfb 1119 crossing} and \textsc{sfb 1053 maki}, and
the \textsc{bmbf} and the State of Hesse within \textsc{crisp-da}.

\end{acks}

\bibliographystyle{ACM-Reference-Format}
\bibliography{bibliographies} 

\end{document}